# Differential formulation of Schrödinger equation leads to vanishing Berry phase

Yong Tao[†]

School of Economics and Business Administration, Chongqing University, Chongqing, China

**Abstract:** The Poincare-Hopf theorem states that a globally smooth tangent vector does not exist on a manifold whose Euler characteristic is non-zero. Nevertheless, when one defines a differential equation on such a manifold, this theorem is always ignored. For example, the differential formulation of Schrödinger equation is defined as a form so that a tangent vector is a product of Hamiltonian and state vector. In this case, if the Hamiltonian and the state vector are both globally smooth functions on parameter space, then the tangent vector will be compelled to become smooth so that the Euler characteristic of the parameter space must be zero. As a result, some Berry phases related to non-zero Euler characteristic will be ruled out.

## 1. Introduction

In general, a sensible physical law can be expressed as an equation; mathematically, this equation may be written as a differential formulation or an integral formulation. This means that there are two ways of describing physical law: differential equation and integral equation. In particular, since Maxwell wrote down the differential formulation of the electromagnetic field equation, it was widely accepted that the differential equation is a logical way of completely describing physical reality, and that there is no physical difference between a differential equation and its integral formulation. This is the reason why Einstein believed that a complete field model may be on the basis of some differential equations (we call differential equation the local description). From that time onward, many physicists were convinced that all physical phenomenons can be understood within the framework of local description. However, when AB phase was discovered in1960s [1], physicists surprisingly noticed that there may exist some physical phenomenons which can not be understood well in the framework of local description. For example, the AB phase is path-dependent, but the path-dependent behaviour is essentially related to some integral equations (we call integral equation the global description). Later, Yang [2] developed the integral formulation for gauge field and further pointed out that what provides an intrinsic and complete description of electro-magnetism is such an integral formulation. As an application of this notion, Wu and Yang proved that the magnetic monopole is self-contained only in the framework of integral formulation [3].

Not only that, recently, the validity of the application of the quantum adiabatic theory (QAT) had been doubted [4-11]. In particular, Pati and Rajagopal [6] showed that the traditional adiabatic condition, $\langle n(t)|\frac{\partial}{\partial t}|m(t)\rangle = 0$ $(n \neq m)$ [12], may result in vanishing Berry phase. More recently, we [13-14] pointed out that such a condition has nothing to do with the adiabatic process, and instead is a consequence of misusing the differential formulation of Schrödinger equation (DFSE) [15]. Our concrete study [14] showed that there is no any inconsistency if the QAT is derived using the integral formulation of Schrödinger equation (IFSE). Conversely, if we use the DFSE to derive the QAT, then the Berry phase would vanish [13-14]. This means that there exists some irreconcilable conflict between Berry phase and DFSE. Consider that Berry phase has been observed by many experiments and also that the role of it in the study of quantum mechanical system spans a vast array of fields, including induced gauge field [16], adiabatic quantum computation [17] and spin Hall effect [18-19], etc., we must make clear the origin of this conflict. If we note that AB phase, magnetic monopole and Berry phase are, without exception, topological phenomenons in physical system, we shall have to infer whether or not the structure of a differential equation itself determines that it fails to describe some topological phenomenons. This is a natural thought, because a differential equation itself presents a differential structure but we can not guarantee that such a differential structure has to be compatible (self-contained) with the topological structure of the corresponding physical system. More concretely, we explore a general type of differential equation of time evolution in the form:

$$\frac{\partial}{\partial t}\mathbf{X}(R) = \mathbf{f}[t, \mathbf{X}(R)], \qquad (1)$$

where the vector $\mathbf{X}(R)$ denotes state variable and $R = R(t)$ denotes physical parameter depending on time variable $t$.

---

[†] Correspondence Email: taoyingyong2007@yahoo.com.cn

As such, our question is as follow:

If $\mathbf{f}[t, \mathbf{X}(R)]$ is a smooth function of $R$, does the differential equation (1) always hold?

The purpose of this paper is to show that the differential equation (1) is *ill-defined* when the *Euler characteristic* of the parameter space (manifold) consisting of all $R$ is non-zero. In particular, to guarantee that the differential equation (1) is *well-defined*, some topological phenomenons (e.g. Berry phase or anholonomy angle) related to non-zero Euler characteristic will be ruled out. Because the differential equation (1) has covered a large class of physical equations, e.g. Maxwell equation, Schrödinger equation and so on, our conclusion will have universal meaning in physics.

## 2. Berry phase related to non-zero Euler characteristic

To see the physical meaning of our study more clearly, this paper only investigates the DFSE. Although the DFSE is a special case of the differential equation (1), we need to emphasize that our main result also holds for the differential equation (1).

The DFSE is written in the form [14,20]:

$$i\frac{\partial}{\partial s}\psi_T(s) = TH(s)\psi_T(s), \qquad s \in [0,1] \qquad (2)$$

where, $T$ denotes the total evolution time and $s = \frac{t}{T}$.

If a quantum system is not isolated from its environment, the observables are described by operators that depend on a set of parameters, $R = \{R^j(s)\}_{j=1}^{l}$. For this quantum system, the parameters $R$ label the points of a smooth manifold $M$, which is called the parameter space of the quantum system. In particular, the Hamiltonian operator $H(R)$ is a smooth and single-valued function of $R \in M$ [12,20]. Here by the smoothness of the Hamiltonian we mean that the state vector $\psi_T(R)$ is a smooth function of $R \in M$, too [21].

In such a parameter space, the DFSE (2) can be rewritten in the form:

$$i\frac{\partial}{\partial s}\psi_T(R) = TH(R)\psi_T(R) \qquad (3)$$

Strictly speaking, according to the mathematical analysis theorem [22], the derivative $\frac{\partial}{\partial s}\psi_T(R)$ exists if and only if the following two conditions holds:

(a). $\psi_T(R)$ is differentiable with respect to $R$.

(b). $R = R(s)$ is differentiable with respect to $s$.

In particular, when (a) and (b) hold, the derivative $\frac{\partial}{\partial s}\psi_T(R)$ is determined by a chain rule [22]:

$$\frac{\partial}{\partial s}\psi_T(R) = \sum_{j=1}^{l}\frac{\partial}{\partial R^j}\psi_T(R)\cdot\frac{dR^j}{ds}. \qquad (4)$$

Using the equation (4), an equivalent form of the DFSE (3) yields:

$$i\sum_{j=1}^{l}\frac{\partial}{\partial R^j}\psi_T(R)\cdot\frac{dR^j}{ds} = TH(R)\psi_T(R). \qquad (5)$$

Notably, (a) may not hold whenever $T \to \infty$ so that the equation (3)-(5) are invalid. However, the adiabatic limit $T \to \infty$ is merely an ideal situation which, in reality, does not occur [23]. For example, many experiments confirm that the existence of Berry phase are carried out for finite but large $T$, not infinite $T$. Because of this, the equation (3)-(5) always hold in reality without worrying about the question of adiabatic limit. Strictly speaking, the existence of Berry phase does not depend on the adiabatic limit. To see this, we survey the following well known example:

***Example 1***: The Hamiltonian of a spin-half particle in a rotating magnetic field is written in the form [13]:

$$\begin{aligned}H(R) &= \mu(R_x(s) \quad R_y(s) \quad R_z(s))\cdot(\sigma_x \quad \sigma_y \quad \sigma_z) \\ &= \mu(B\sin\theta\cos 2\pi s \quad B\sin\theta\sin 2\pi s \quad B\cos\theta)\cdot(\sigma_x \quad \sigma_y \quad \sigma_z)\end{aligned} \qquad (6)$$

where $s \in [0,1]$.

The Hamiltonian (6) of course satisfies (a) and (b). As is well known, the parameter space of this physical system is a two-dimensional spherical surface $S^2$, that is,

$$S^2 = \{(R_x(s)\ R_y(s)\ R_z(s)) \mid R_x(s)^2 + R_y(s)^2 + R_z(s)^2 = B^2\}. \quad (7)$$

In particular, the Berry phase of this physical system, which has nothing to do with adiabatic limit, is merely determined by the anholonomy angle of the spherical surface $S^2$. For example, if a state vector transports around a spherical triangle $\Delta ABC$ on $S^2$ (sees Fig. 1), then it will acquire a Berry phase $\gamma(\Delta ABC)$ which is equal to the half of the solid angle $\Omega(\Delta ABC)$ (anholonomy angle) subtended by the spherical triangle $\Delta ABC$; that is [24],

$$\gamma(\Delta ABC) = \pm \frac{1}{2} \Omega(\Delta ABC). \quad (8)$$

Not only that, consider that the solid angle of the whole sphere $S^2$ equals $2\pi\chi(S^2)$ [25], the solid angle $\Omega(\Delta ABC)$ will be denoted by the following formula:

$$\Omega(\Delta ABC) = \frac{\Pi(\Delta ABC)}{\Pi(S^2)} \cdot 2\pi\chi(S^2), \quad (9)$$

where $\Pi(\Delta ABC)$ denotes the area of $\Delta ABC$, $\Pi(S^2)$ denotes the area of $S^2$ and $\chi(S^2)$ denotes the *Euler characteristic* of $S^2$.

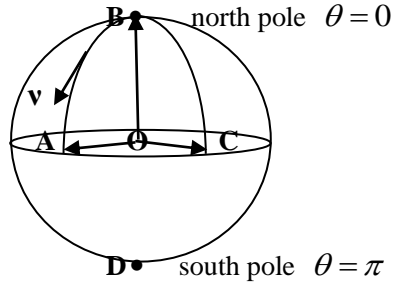

**Figure. 1:** A state vector $\mathbf{v}$ transports around a spherical triangle $\Delta ABC$ on $S^2$, then it will acquire a Berry phase $\gamma(\Delta ABC)$ which is equal to the half of the solid angle $\Omega(\Delta ABC)$ (anholonomy angle) subtended by the spherical triangle $\Delta ABC$.

The equations (8) and (9) together indicate that there is a relationship between Euler characteristic and Berry phase as follow:

$$\gamma(\Delta ABC) = \pm \frac{\Pi(\Delta ABC)}{\Pi(S^2)} \cdot \pi \chi(S^2). \tag{10}$$

## 3. Application of Poincare-Hopf theorem

The equation (10) undoubtedly implies that for *some* parameter space (e.g. $S^2$) the existence of Berry phase depends on non-zero Euler characteristic of this parameter space. However, next, we shall prove a theorem which states that the DFSE (5) holds everywhere on $S^2$ if and only if $\chi(S^2) = 0$. To arrive at this theorem, we first introduce an important lemma-namely, the *Poincare-Hopf theorem*.

*Lemma 1* (**Poincare-Hopf theorem**) [26]**:** Let $X$ be a smooth tangent vector field on a compact manifold $M$. If $X$ has only isolated zeros (singularities), then

$$Index(X) = \chi(M), \tag{11}$$

where, $Index(X)$ denotes the total index of isolated zeros of $X$.

Here we need to introduce the mathematical definition of isolated zero.

*Definition of isolated zero*: $p^j$ $(j=1,\ldots,l)$ is an isolated zero of $X$ in the manifold $M$ if and only if $X(p^j) = 0$ and for every point $q^j \in U_\varepsilon(p^j) - \{p^j\}$ there exists a positive number $\delta$ such that $|X(q^j)| \geq \delta$, where $U_\varepsilon(p^j) - \{p^j\}$ denotes some neighborhood [22] of $p^j$ which does not include $p^j$.

For example, the two-dimensional spherical surface $M = S^2$ has two isolated zeros, which are north and south poles respectively, sees the points B and D in Fig. 1. Because of this, by lemma 1 we have $\chi(S^2) = 2$. Thus, the solid angle of the whole sphere $S^2$ equals $4\pi$. More importantly, the lemma 1 strongly indicates that if $\chi(M) \neq 0$, then there at least exists an isolated zero $p^j$ in the manifold $M$ so that the tangent vector $X(p^j) = 0$.

With these preparations, we proceed to prove the main result of this paper.

**Theorem 1:** If the Hamiltonian $H(R)$ is a *smooth function* of $R \in M$ and if the DFSE (5) holds everywhere on the manifold $M$, then we have

$$\chi(M) = 0.$$

*Proof.* Let us assume that DFSE (5) holds everywhere on the manifold $M$ and that $\chi(M) \neq 0$.

Because $H(R)$ and $\psi_T(R)$ are smooth functions of $R \in M$, comparing both sides of the DFSE (5) we can conclude that $\frac{\partial}{\partial R^j}\psi_T(R)$ is a continuous (smooth) function of $R \in M$, too. Because of this, for any two points $x^j$ and $y^j$ in the manifold $M$ such that $|x^j - y^j| \to 0$, there does have [22]

$$\left|\frac{\partial}{\partial R^j}\psi_T(x^j) - \frac{\partial}{\partial R^j}\psi_T(y^j)\right| \to 0. \qquad (12)$$

On the other hand, $\frac{\partial}{\partial R^j}\psi_T(R)$ is obviously a tangent vector on the manifold $M$. Because $\chi(M) \neq 0$, by the lemma 1 there at least exists an isolated zero $p^j$ in the manifold $M$ so that the tangent vector $\frac{\partial}{\partial R^j}\psi_T(p^j) = 0$ and for every point $q^j \in U_\varepsilon(p^j) - \{p^j\}$ there has $\left|\frac{\partial}{\partial R^j}\psi_T(q^j)\right| \geq \delta$. This means that we have $|p^j - q^j| \to 0$ so that

$$\left|\frac{\partial}{\partial R^j}\psi_T(p^j) - \frac{\partial}{\partial R^j}\psi_T(q^j)\right| \geq \delta.$$

This result is clearly contrary to the equation (12). To remove this contradiction, we do have

$$\chi(M) = 0.$$

The proof is complete. □

As such, by the theorem 1 the DFSE (5) holds everywhere on $S^2$ if and only if $\chi(S^2) = 0$. Notably, if $\chi(S^2) = 0$, then the equation (10) implies that Berry phase vanishes[1]! So, for the physical system specified by the example 1, the DFSE (5) does lead to vanishing Berry phase. This result is exhibiting the shortcoming of differential equation in studying the global property of some geometrical manifold.

---

[1] This is a natural result if we remember that many geometrical phases are due to non-zero Chern number [27-28]. Consider that the Euler characteristic is a *special Chern number* (accurately, called the top Chern class of the base manifold) [29], we can understand why zero Euler characteristic will rule out some (*not all*) Berry phases (e.g. Example 1).

## 4. Shortcoming of differential description

The theorem 1 has confirmed that the DFSE (5) does lead to vanishing Berry phase in the case of example 1. In fact, the mathematical origin of such a result can be found in global differential geometry. The global differential geometry asserts that the derivative of a state vector (tangent vector) can not be defined on a global manifold which is globally non-trivial (for example, the Euler characteristic of this manifold is non-zero). Generally speaking, the derivative of a state vector may be defined on different coordinate patches which together constitute the global manifold. This is a shortcoming of differential description. However, Hamiltonian and state vector may be globally smooth, and can be therefore defined on a global manifold, e.g. the example 1. Unfortunately, t*he DFSE compels the derivative of the state vector to equal the product of Hamiltonian and state vector*. From this reasoning, the derivative of the state vector shall be also well defined on the global manifold (i.e., $S^2$) in the example 1. This clearly contradicts the assertion of global differential geometry since the Euler characteristic of $S^2$ is non-zero.

Notably, the smoothness of Hamiltonian $H(R)$ emphasized by theorem 1 is very important. For example, because of this, the DFSE (5) can not rule out the AB phase. To see this, we only need to observe that, in the Hamiltonian $H(R)=\frac{1}{2}(-i\nabla_R - A_R)^2$ specified by AB effect [1], there exists a singular (no smooth) function (vector potential) $A_R$ which will guarantee that $H(R)$ is not smooth so that the theorem 1 is invalid. In particular, in the framework of differential formulation, we can not remove $A_R$ from $H(R)$ via the local gauge transformation $\tilde{A}_R = A_R - \nabla_R \Lambda$. This is because we can remove $A_R$ from $H(R)$ if and only if $A_R = \nabla_R \Lambda$ where $\Lambda$ is a globally smooth function of $R$ (otherwise, the field strength $F$ is not gauge invariant). Unfortunately, the AB phase must be zero if $A_R = \nabla_R \Lambda$.

Interestingly, although we can not remove $A_R$ from $H(R)$ via the gauge transformation, we may modify the domain of parameter space so as to guarantee that $H(R)$ is a smooth function on the new parameter space. For example, if the parameter space of AB effect is denoted by the plane $M = R^2$ and if the position occupied by the flux is denoted by the point $O$, then $H(R)$ will become a smooth function of $R$ on the modified parameter space $\tilde{M} = R^2 - \{O\}$.

If the AB phase is due to Euler characteristic of $\tilde{M}$, then, according to the theorem 1, the DFSE (5) will rule out AB phase. However, this case will not occur, it is because the AB phase is essentially due to other Chern number related to (fiber bundle of) $\tilde{M}$ (hole effect [27]). Of course, consider that many geometrical phases are due to non-zero Chern number [27-28], a natural thought is to generalize the theorem 1 to the general case where the Euler characteristic is replaced by the Chern number. However, we shall discuss this in the future work.

## 5. Adiabatic limit

In the other direction, someone perhaps argue that the theorem 1 is on the basis of non-adiabatic limit ($T < \infty$), and thereby may be invalid in the adiabatic limit $T \to \infty$. However, next we shall show that the DFSE, in the adiabatic limit $T \to \infty$, still leads to vanishing Berry phase. To this end, we express the state vector $\psi_T(s)$ in the basis $\{|n(s)\rangle\}$ [5],

$$\psi_T(s) = \sum_n a_n(s,T) \exp\left[-iT\int_0^s E_n(s')ds'\right]|n(s)\rangle. \qquad (13)$$

Substituting equation (13) into DFSE (2) yields

$$\frac{\partial}{\partial s}a_m(s,T) = -a_m(s,T)\langle m(s)|\frac{\partial}{\partial s}|m(s)\rangle$$
$$- \sum_{n \neq m} a_n(s,T) \exp\left[iT\int_0^s (E_m(\sigma) - E_n(\sigma))d\sigma\right]\langle m(s)|\frac{\partial}{\partial s}|n(s)\rangle. \qquad (14)$$

In the rotating representation, the QAT is written as $\lim_{T \to \infty} a_m(s,T)$, which converges uniformly to a continuous function $a_m(s)$ [5]. Clearly, the QAT $a_m(s)$ is a solution of the DFSE (14) as $T \to \infty$ if and only if

$$\frac{\partial}{\partial s}a_m(s) = -a_m(s)\langle m(s)|\frac{\partial}{\partial s}|m(s)\rangle$$
$$- \sum_{n \neq m} a_n(s) \cdot \left\{\lim_{T \to \infty} \exp\left[iT\int_0^s (E_m(\sigma) - E_n(\sigma))d\sigma\right]\right\} \cdot \langle m(s)|\frac{\partial}{\partial s}|n(s)\rangle. \qquad (15)$$

It is easy to see that the left-hand side of equation (15) does not depend on $T$, but that the right-hand side depends on $T$. This is a clear contradiction. To rule out this contradiction, we must demand:

$$\langle n(s)|\frac{\partial}{\partial s}|m(s)\rangle = 0. \quad (n \neq m) \qquad (16)$$

Substituting equation (16) into equation (15) yields

$$\begin{cases} \dfrac{\partial}{\partial s}a_m(s) = -a_m(s)\langle m(s)|\dfrac{\partial}{\partial s}|m(s)\rangle \\ \langle m(s)|\dfrac{\partial}{\partial s}|n(s)\rangle = 0 \quad (m \neq n) \end{cases} \tag{17}$$

Integration of the first equation of the equations (17) gives

$$\begin{cases} a_m(s) = \exp\left[-\int_0^s \langle m(\sigma)|\dfrac{\partial}{\partial \sigma}|m(\sigma)\rangle d\sigma\right] \\ \langle m(s)|\dfrac{\partial}{\partial s}|n(s)\rangle = 0 \quad (m \neq n) \end{cases} \tag{18}$$

Obviously, the first equation of the equations (18) gives Berry phase factor. Notably, there is a constraint (i.e., equation (16)) aimed to the Berry phase factor[2]. Unfortunately, we can not abandon this constraint. This is because $\dfrac{\partial}{\partial s}a_m(s) = -a_m(s)\langle m(s)|\dfrac{\partial}{\partial s}|m(s)\rangle$ holds if and only if the equation (16) holds. In fact, the existence of this constraint is just the reason why the DFSE leads to vanishing Berry phase. To see this, we turn to the example 1.

The eigenstates of the Hamitonian (6) are as follows [13]:

$$|\uparrow(s)\rangle = \begin{pmatrix} \cos\dfrac{\theta}{2} \\ \sin\dfrac{\theta}{2}\exp(i2\pi s) \end{pmatrix}, \tag{19}$$

$$|\downarrow(s)\rangle = \begin{pmatrix} -\sin\dfrac{\theta}{2}\exp(-i2\pi s) \\ \cos\dfrac{\theta}{2} \end{pmatrix}, \tag{20}$$

where $\theta$ is an arbitrary real number in interval $[0, \pi]$.

Substituting equations (19) and (20) into equations (18) yield

$$\begin{cases} a_m(1) = \exp[i\gamma(C)] = \exp[-i\pi(1-\cos\theta)] \\ \theta = 0 \quad or \quad \pi \end{cases} \tag{21}$$

Clearly, the equations (21) give $\exp[i\gamma(C)] = 1$; that is, the Berry phase vanishes.

As such, we indeed see that the DFSE in the adiabatic limit $T \to \infty$ leads to vanishing Berry phase. Although we only investigate the example 1 here, this is a general conclusion [13]. As seen in the example 1, the DFSE, in essence, imposes a special constraint (equation (16)) on the parameter space $S^2$. Because of this constraint, $S^2$ will degenerate into a point ($\theta = 0$ or $\pi$) so that the Berry phase vanishes (sees Fig. 1). This is consistent with the conclusion of the theorem 1. In fact, north and south poles ($\theta = 0$ or $\pi$) are just two isolated zeros of $S^2$ which induce $\chi(S^2) = 2$. However, $\chi(S^2) = 2$ is irreconcilable with the DFSE. Mathematically, the DFSE essentially presents a differential structure, but such a differential structure is incompatible with the topological structure of $S^2$. If we insist on imposing this differential structure on $S^2$, then the topological structure of $S^2$ will be destroyed so that the Berry phase (or anholonomy angle) vanishes. This result is exhibiting the spirit of Milnor's 7-sphere [30]; that is, *the differential structure may be independent of the topological structure*. Consider that some Berry phase is a outcome of topological structure, it may be of course ruled out by a differential structure.

---

[2] In the past, such a constraint is always ignored.

## 6. Conclusion

The differential formulation of Schrödinger equation essentially presents a differential structure. Consider that the differential structure may be independent of the topological structure, we can not arbitrarily impose a differential structure on a topological space. In particular, if we impose the differential formulation of Schrödinger equation on some physical system, and meanwhile if the Euler characteristic of parameter space of this physical system is non-zero, then the topological structure of the parameter space will be destroyed. As a result, some topological phenomenon (e.g. Berry phase or anholonomy angle) may vanish. This reminds us that the differential equation is not a logical way of completely describing physical reality. For example, it may destroy the topological structure of physical system and thereby fail to describe some topological phenomenon. To avoid this difficulty, we suggest to use the integral formulation of corresponding differential equation.